\documentclass[a4paper,11pt]{article}
\usepackage{jinstpub} 
\usepackage{lineno}
\usepackage{siunitx}

\title{\boldmath First generation 4H-SiC LGAD production and its performance evaluation}

\author{R. Novotný$^{a}$, J. Chochol$^b$, V. Kafka$^a$, A. Klimsza$^b$, A. Kozelsky$^b$, J. Kroll$^c$, R. Malousek$^b$, M. Marčišovský$^a$, M. Marčišovská$^a$, M. Mikeštíková$^c$, D. Novák$^b$, P. Slovák$^b$, R. Špetík$^b$, P. Švihra$^{a,c}$}
\affiliation{$^a$Faculty of Nuclear Sciences and Physical Engineering, Czech Technical University in Prague,\\
Břehová 7, 115 19 Prague 1, Czech Republic}
\affiliation{$^b$onsemi - ON Semiconductor Czech Republic,\\
Rožnov p.R., 1. máje 2230, 756 61, Czech Republic}
\affiliation{$^c$Institute of Physics of the Czech Academy of Sciences,\\
Na Slovance 1999/2, 182 21 Prague 8, Czech Republic}

\emailAdd{radek.novotny@fjfi.cvut.cz}

\abstract{This contribution will delve into the design and performance of the newly produced Silicon Carbide Low Gain Avalanche Detectors (4H-SiC LGADs) and provide a comprehensive summary of their measured characteristics. This includes an analysis of the detector's performance, temperature stability, and the effectiveness of the internal gain layer in improving signal generation.

The 4H-SiC is re-emerging as a strong candidate for the next generation of semiconductor detectors. This material offers several advantages, including high radiation tolerance and the ability to operate over a wide range of temperatures without significant annealing effects. However, the signals generated by minimum ionizing particles in the 4H-SiC detector are lower compared to the signal produced by standard silicon detectors due to their higher bandgap energy. This is addressed by implementing a charge multiplication layer, which results in the intrinsic gain of the device.

The presented 4H-SiC LGADs produced by onsemi are specifically designed and optimized for fabrication on the n-type substrate/epi wafer with the gain layer implanted approximately $1~\mathrm{\mu m}$ below the surface. The first iteration of these LGAD structures was manufactured in early 2024 and since then has been subjected to laboratory evaluation. The measured properties of these detectors align well with the predictions arising from the extensive TCAD simulation studies.}

\keywords{Detector design and construction technologies and materials, Radiation-hard detectors, Solid state detectors}

\begin{document}
\maketitle
\flushbottom

\section{Introduction}
\label{sec:intro}
Recent advancements in the manufacturing processes of wide bandgap semiconductor materials have revitalized interest in high-energy detectors based on these materials. Among these, Silicon Carbide (SiC) has emerged as a leading candidate due to its exceptional properties, including high radiation tolerance, excellent thermal stability, and the ability to operate in harsh environments. As the demand for advanced detectors in high-energy physics and other applications increases, the advantages offered by SiC make it an attractive alternative to traditional silicon-based detectors.

4H-SiC, in particular, has gained attention for its superior performance characteristics. With a wider bandgap energy than silicon, 4H-SiC exhibits lower intrinsic noise levels and higher breakdown voltages, making it suitable for applications that require robust performance in high-radiation fields \cite{SiC-applications}. However, the lower charge generated in the bulk material due to its wider bandgap presents a challenge. To address this limitation, the implementation of a charge multiplication layer is a preferred design choice, resulting in the development of Low Gain Avalanche Diodes (LGADs) \cite{LGAD}. This technology has shown promising results in standard silicon detectors and is now being adapted for SiC detectors.

This paper is focused on the design, fabrication, and performance evaluation of the first-generation 4H-SiC LGADs produced by onsemi. These sensors are specifically engineered to enhance signal generation while leveraging the high radiation tolerance inherent to SiC. Extensive laboratory evaluations have been conducted to assess their performance, temperature stability, and the effectiveness of their design in signal amplification, with initial results aligning closely with TCAD simulation predictions. This contribution aims to summarize the characteristics of the newly produced 4H-SiC LGADs, demonstrating their potential applications in high-energy physics and particle detection in general.

\section{Sensor design and fabrication}
\label{sec:design}

\begin{figure}[!htp]
    \centering
    \includegraphics[width=0.49\linewidth]{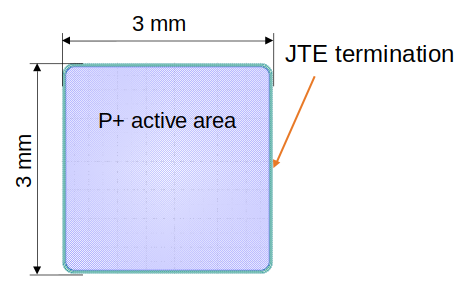}
    \includegraphics[width=0.49\linewidth]{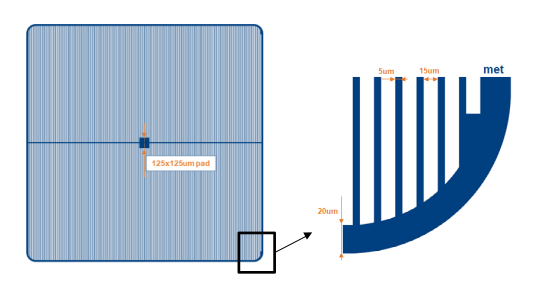}
    \caption{Schematic representation of the sensor design. Dimensions of the sensor layout (left). Visualization of the metal mask with a grid structure enabling optical measurements (right).}
    \label{fig:MetalMask}
\end{figure}

The sensors are designed as $3 \times 3 \, \si{\milli\meter\squared}$ squares with a top layer either fully covered by metal or by the metal grid with \SI{5}{\micro\meter} wide stripes and \SI{15}{\micro\meter} pitch between these stripes enabling characterization of samples with the light source.
 For reference measurements, each production wafer features LGAD sensors with an internal gain layer and PN diodes with no internal gain.
The sensor layout with visualization of the top metalization is shown in Figure~\ref{fig:MetalMask}.

\begin{figure}[!htp]
    \centering
    \includegraphics[width=0.8\linewidth]{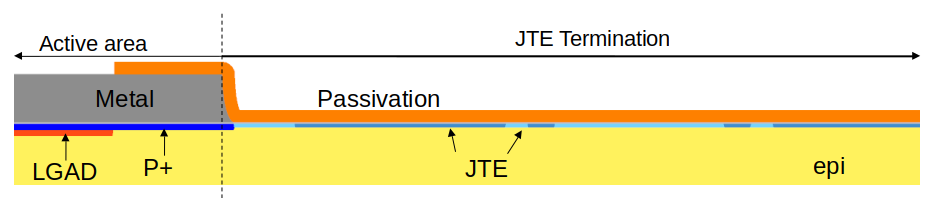}
    \caption{Illustration of the full 4H-SiC LGAD detector design, including the epitaxial layer, gain layer, and junction termination extension for a reasonably high breakdown voltage.}
    \label{fig:DetectorLayout}
\end{figure}

The substrate is a standard commercially available $6^{\prime\prime}$ n-type 4H-SiC wafer with doping concentration $\sim$~\SI{e19}{\per\cubic\cm}. On the top of the substrate wafer, the high resistivity n-type epitaxial layer with the thickness between 30 and \SI{100}{\micro\meter} and doping concentration down to $\sim$ \SI{5e13}{\per\cubic\cm} is grown.
The epitaxial growth is followed by the manufacturing of Junction Termination Extension (JTE) and $P^+$ areas which are created by $P^+$ chain implants with energies in the range of \SIrange[range-phrase = -, range-units = single]{30}{200}{\kilo\eV} and doses in range of \SIrange[range-phrase = -, range-units = single]{1}{8e14}{\per\square\cm}. 
The JTE implantation used at the edge of the sensor is designed for keeping the breakdown voltage above \SI{1}{\kilo\volt}. 
The LGAD implant, which particular implementation is highly dependent on designed gain, is a single n-type implant with energy in range \SIrange[range-phrase = -, range-units = single]{950}{1250}{\kilo\eV} and doses reaching \SIrange[range-phrase = -, range-units = single]{1.5}{1.8e13}{\per\square\cm}. 
The top contact is ensured by the thick metal layer. 
The passivation layer is placed at the edge of the sensor.
The full detector design is shown in Figure~\ref{fig:DetectorLayout}.

\begin{figure}[!htp]
    \centering
    \includegraphics[width=0.9\linewidth]{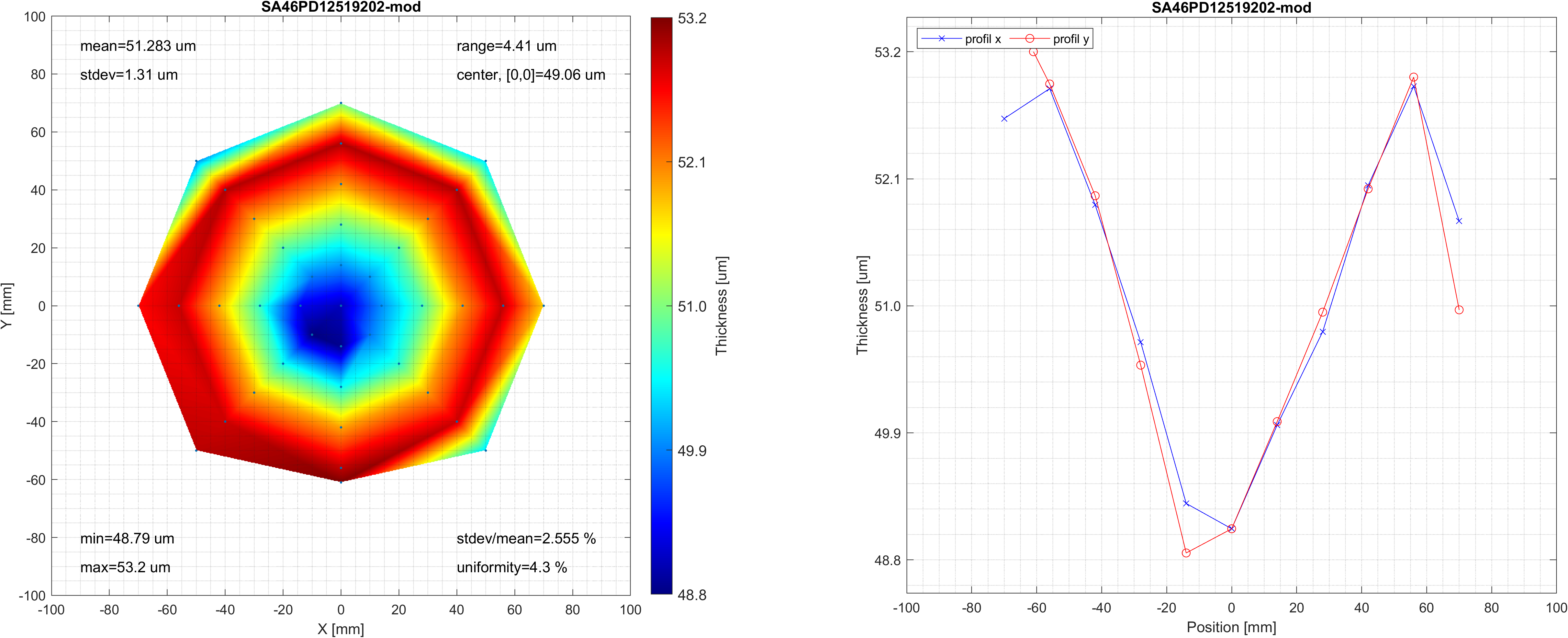}
    \caption{FTIR thickness map of a wafer with a \SI{50}{\micro\meter} epitaxial layer, showing thickness uniformity across the wafer. }
    \label{fig:FTIR}
\end{figure}

The designed LGAD sensors were entirely fabricated at onsemi foundries in Rožnov pod Radhoštěm, ensuring precise control over every step of the production process.
Achieving thick epitaxial growth with a doping concentration below \SI{1e14}{\per\cubic\cm} presents significant challenges, requiring meticulous control of the growth parameters. 
The epitaxial layer thickness was measured using Fourier Transform Infrared Spectroscopy (FTIR), a precise and non-destructive technique \cite{FTIR}.
Figure \ref{fig:FTIR} illustrates an epitaxial thickness map of a wafer with a \SI{50}{\micro\meter}  thick epitaxial layer. The layer's thickness tends to increase toward the wafer edges, but the achieved uniformity of $4.3\%$ remains well within acceptable limits for reliable device fabrication. 
This performance is especially notable given the inherent difficulty of producing high-quality thick SiC epitaxial layers.

\section{Device testing}
\label{sec:test}

\begin{figure}[!htp]
    \centering
    \includegraphics[width=0.47\linewidth]{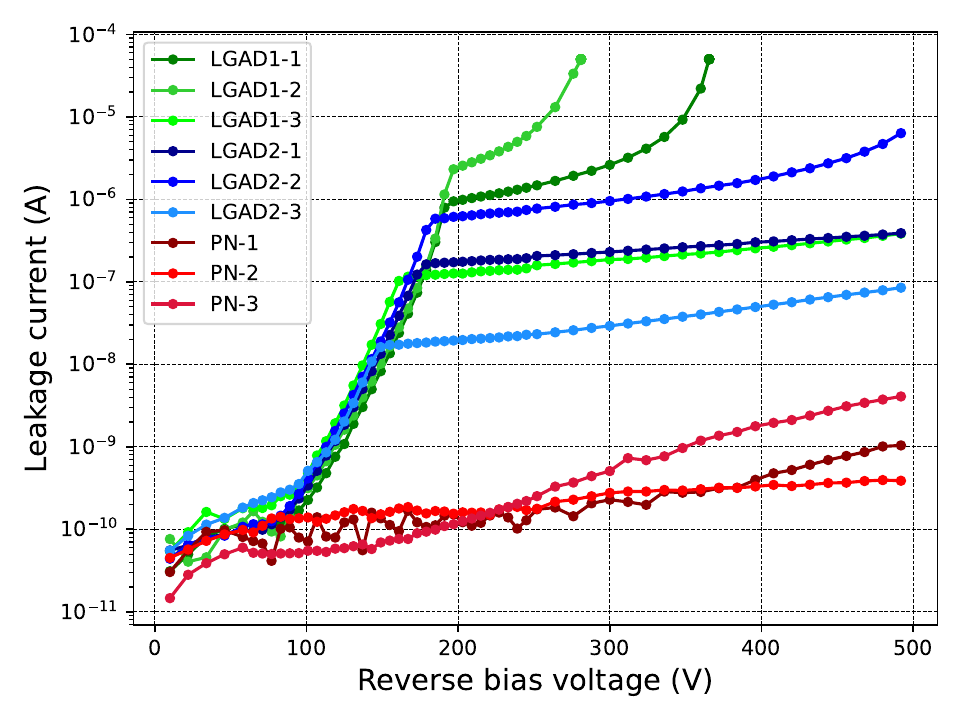}
    \includegraphics[width=0.47\linewidth]{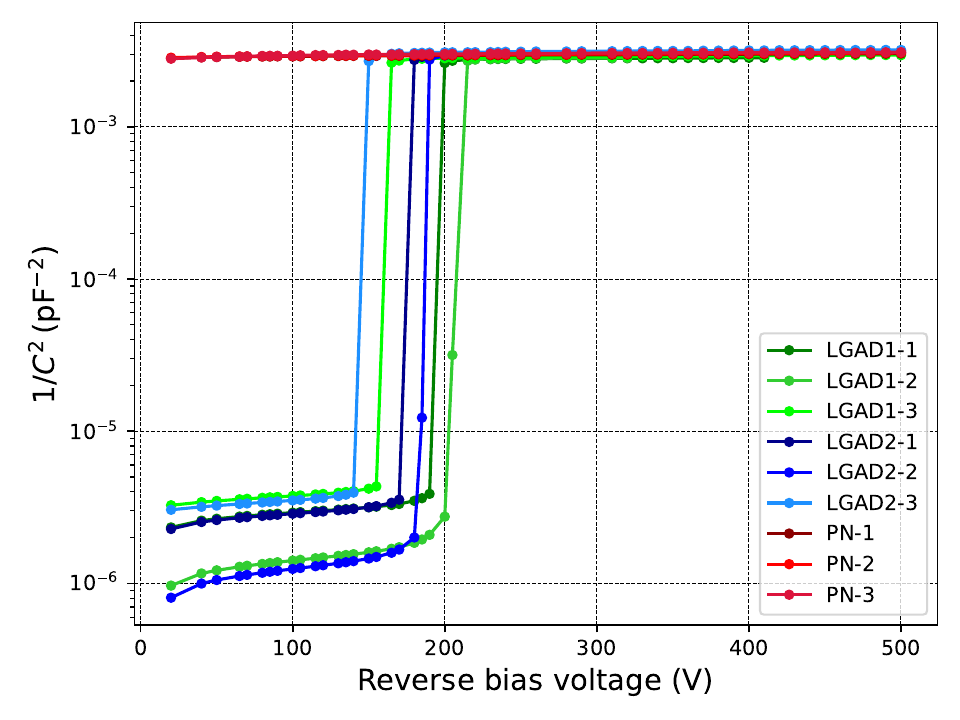}
    \caption{Measured IV (left) and CV (right) curves of the manufactured sensors from the 3 wafers with \SI{30}{\um} thick epitaxial layer and different parameters of epitaxial implantation and gain layer. The depletion of the gain layer occurs between \SI{155}{\volt} and \SI{220}{\volt}, aligning well with TCAD simulation predictions.}
    \label{fig:IVCV}
\end{figure}

The first iteration of manufactured sensors was fully tested at the wafer level at the CTU and FZU institutes using the TS3000SE and TESLA200 probe stations, respectively. 
The first batch consists of three wafers using \SI{30}{\micro\meter} thick epitaxial layer with different implantation strategies and gain layer parameters. Each wafer contains reference PN sensors and two types of LGAD sensors, labeled LGAD1 and LGAD2, which differ in their designed gain. 
The final labeling scheme follows the format \textit{Type - WaferNumber}, e.g., LGAD1-2 or PN-1.
The IV and CV curves were measured at room temperature by applying reverse bias voltage in the range of \SIrange[range-phrase = -, range-units = single]{0}{500}{\V}. 
The IV curves measured for all sensor types are given in Figure~\ref{fig:IVCV}, showing the dependency of the breakdown voltage on the designed gain factor. 
The depletion of the gain layer occurs in the range of \SIrange[range-phrase = -, range-units = single]{155}{220}{\V}, aligning well with TCAD simulations; however, the full depletion voltage could not be precisely measured at the wafer level.
The capacity of the fully depleted sensor is $\SI[separate-uncertainty=true]{18.0\pm 0.2}{\pico\farad}$ which is also in alignment with expected values.

\begin{figure}[!htp]
    \centering
    \includegraphics[width=0.47\linewidth]{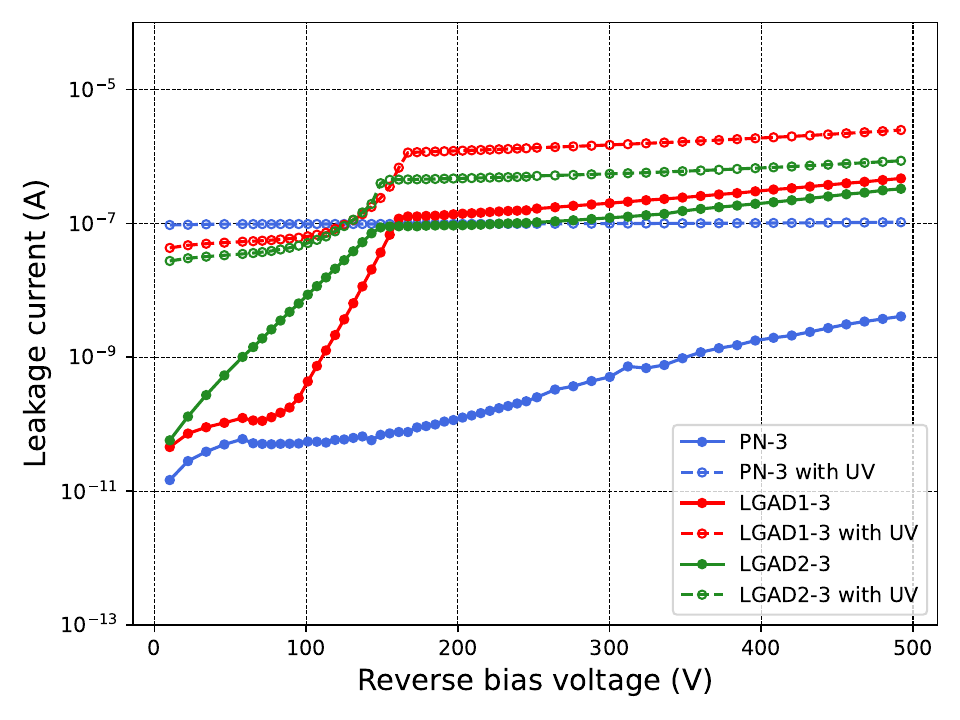}
    \includegraphics[width=0.47\linewidth]{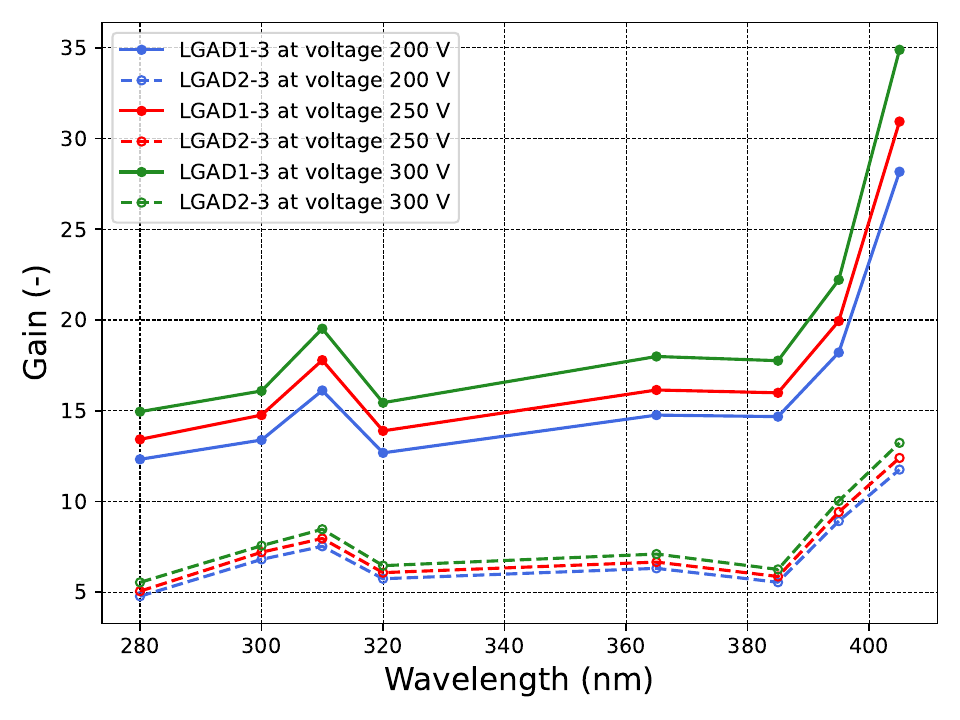}
    \caption{Response to UV light: Leakage current of the sensors measured for the wavelength of \SI{277}{\nm}~(left) and gain as a function of wavelength (right).}
    \label{fig:UV_gain}
\end{figure}

The response of the sensors was tested using multiple UV-LED sources with wavelengths between 277 and \SI{405}{\nm} coupled to the optical fiber. 
The fiber was placed roughly \SI{5}{\mm} above the center of the sensor which allows us to measure response without the risk of parasitic effects caused by UV light passing through the passivation layer at the edge of the sensor. 
An example of the sensor response measured with the single wavelength, along with the dependence of the gain on the wavelength - calculated as a ratio between the reverse leakage current measured for the LGAD sensor and the PN diode at a given voltage, is shown in Figure~\ref{fig:UV_gain}.
The gain remains stable up to \SI{380}{\nano\meter}, roughly corresponding to the bandgap energy. Beyond this threshold, two-photon absorption begins to have an effect \cite{TPA}, resulting in the gain increase. 
To accurately estimate its effect on gain, a more detailed study of two-photon absorption with precise UV-light focusing is required.
The measured gain values in the single-photon absorption region align with the simulations and are utilized to refine the charge amplification calculations within the simulations.

\begin{figure}[!htp]
    \centering
    \includegraphics[width=0.47\linewidth]{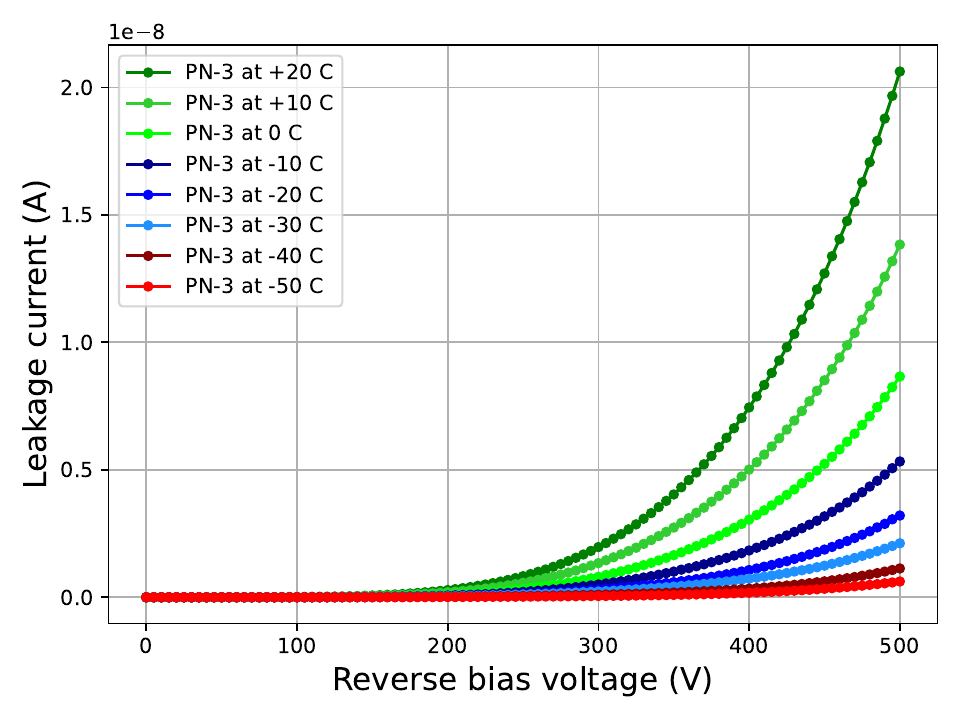}
    \includegraphics[width=0.47\linewidth]{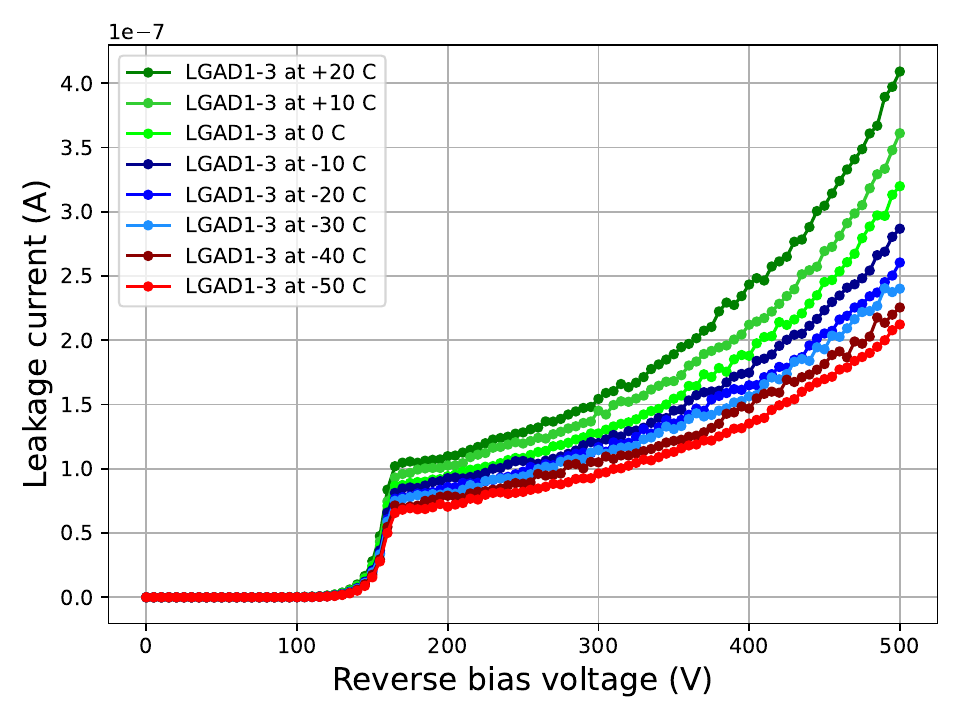}
    \caption{Measured temperature dependence of the PN diode (left) and LGAD1 sensor (right).}
    \label{fig:temperature}
\end{figure}

The thermal stability of the samples was studied in the nitrogen atmosphere of the TESLA200 probe station with the chuck temperature set between \SI{-50}{\degreeCelsius} and \SI{20}{\degreeCelsius}. The IV characteristics of the PN diode and the LGAD sensors are presented in Figure~\ref{fig:temperature}. 
The results reveal that the reverse leakage current of the PN diode doubles approximately every $\SI{15}{\degreeCelsius}$, indicating a slower temperature increase of the leakage current compared to standard silicon devices, which highlights improved thermal stability.

\section{Conclusions}
\label{sec:test}
This study presents the design, fabrication, and performance evaluation of the first-generation 4H-SiC LGADs. The newly developed sensors demonstrate promising characteristics, aligning well with TCAD simulation predictions and meeting the design objectives for high-energy physics applications. The integration of a gain layer successfully compensates for the lower charge generation associated with the wide bandgap of 4H-SiC, resulting in enhanced signal performance. 

Overall, the combination of high radiation tolerance, thermal stability, and customizable design features positions 4H-SiC LGADs as a strong candidate for next-generation radiation detector technologies. 
Future efforts will focus on refining the gain layer design, exploring various device configurations for different segmentation strategies, and extending operational tests to high-radiation environments and real-world applications.
\acknowledgments

This work was supported by the Technological Agency of the Czech Republic - Project TK05020011. 
The team from the Institute of Physics of the Czech Academy of Sciences was supported by OPJAK  FORTE - CZ.02.01.01/00/22\_008/0004632 project.
Researcher Peter Švihra conducts his research under the Marie Skłodowska-Curie Actions – COFUND project, which is co-funded by the European Union (Physics for Future – Grant Agreement No. 101081515).




\end{document}